\documentclass{edp-conf}
\usepackage{graphicx}
\usepackage{floatflt}

\begin{document}

\TitreGlobal{SF2A 2003}

\title{Testing the energy scale of the CELESTE experiment}
\author{Brion, E.}\address{CENBG, Domaine du Haut-Vigneau, BP 120, 33175 Gradignan Cedex, France\\
E-mail: \texttt{brion@cenbg.in2p3.fr}}
\author{Smith, D.$^1$}
\author{the CELESTE collaboration}
\runningtitle{Energy scale of CELESTE}
\setcounter{page}{237}
\index{Brion, E.}
\index{Smith, D.}

\maketitle

\begin{abstract}
The uncertainties on the absolute energy scale of the CELESTE experiment affect the constraints CELESTE can impose on $>30$~GeV $\gamma$-ray emission models. To date, these uncertainties have been of order 30~\%. The optical throughputs of the atmosphere and the detector are key to understanding the energy scale. We are comparing CELESTE photomultiplier anode currents measured for bright stars with predictions from the detector simulation. In these studies we include data from recently installed radiometers, and from the LIDAR operating on the site.
\end{abstract}

%

\section{Introduction}

CELESTE detects the Cherenkov light of the electromagnetic shower produced in the atmosphere by the $\gamma$-rays coming from high energy astrophysical sources. It uses 53 heliostats of the former EDF solar plant in the French Pyrenees at the Th\'emis site. The light is reflected to the secondary optics and photomultipliers (PM) installed at the top of the tower, and is finally sampled to be analysed (Par\'e et al. 2002).

Since 2000, CELESTE has detected the Crab nebula and the blazar Mrk~421 (de Naurois et al. 2002, Manseri et al. 2003) with uncertainties on the energy scale of order 30~\%. A better understanding of the energy scale would better constrain the $\gamma$-ray emission models. These uncertainties are dominated by our knowledge of the atmospheric shower (Monte Carlo simulations) and of the optical throughput. To reduce the uncertainty, we are studying the PM currents for bright stars (\S~\ref{sec:photometry}). We also study the atmosphere with radiometers and a LIDAR (\S~\ref{sec:atmosphere}) that help to determine data quality (CELESTE page for more details).

\section{Photometry} \label{sec:photometry}

Our photometry study focuses on the comparison between simulations and data on bright stars' currents. We take data during clear nights, ON and OFF source, and considering the atmospheric extinction, the ray tracing, the projected heliostat surface, the reflectivity of the mirrors, the properties of the Winston cones (geometry, refraction index), the PM quantum efficiency (relative to 450~nm) and the PM gains we can convert the currents into illuminations:\\
\begin{tabular}{lllll}
$\hspace{-0.2cm}b = \frac{i}{ge}$ & with & $b$: luminosity [p.e./s], && $g$: PM charge gain ($\sim 6\times10^4$),\\
&& $i$: anode currents [A], && $e = 1.6\times10^{-19}$C.\\
\end{tabular}

\begin{floatingfigure}[l]{5.5cm}
   \centering \includegraphics[width=5.5cm]{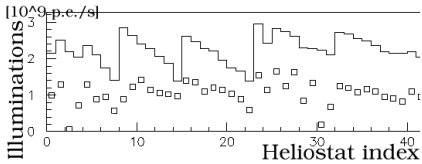}
   \caption{Mu Peg illumination [$10^9$~p.e./s] as a function of the heliostat index for data (points) and simulation (histogram).}
   \label{fig:MuPeg}
\end{floatingfigure}

The simulations of the detector optics predict a higher luminosity than data (for all type of stars, fig.~\ref{fig:MuPeg}). We checked the algorithm at each step of the simulation with the following goals: to compare blue and red stars in the same conditions (because of CELESTE sensitivity of 350 to 550~nm), to have the geometrical differences as a function of hour angle, and to clarify the differences between diffuse and point sources.

Part of the large disagreement between predicted and observed illuminations arises from errors in our gain determination. We are also reviewing the mirror reflectivities and photocathode response. However, we believe that when we conclude a discrepancy will remain. It is likely that \textbf{CELESTE's energy threshold will be revised upwards}, perhaps by 30~\%.

\section{Understanding the atmosphere with radiometers and LIDAR} \label{sec:atmosphere}

Two radiometers are installed at Th\'emis pointing at the zenith and measuring the temperature of the water vapour (cloud or haze) of the sky. They are sensitive to wavelengths of $10\pm3~\mu$m which correspond to the absorption of the earth emission by the water vapour. Seen from the ground, we get the emission of the water vapour (integrated on an unknown distance depending on the column density of the water vapour) plus a part of the earth emission that is absorbed and reemitted. The LIDAR operates in UV (355~nm) and green (532~nm) bands. We aim to determine the atmospheric extinction (opacity) as a function of the altitude (differential measurement), up to $\sim20$~km above the site.

Both instruments are used to \textbf{select good quality data} for CELESTE, the LIDAR by shooting several times per night between the CELESTE runs and the radiometers by taking data simultaneously with the runs.


\end{document}